\documentclass[conference]{IEEEtran}
\usepackage{cite}
\usepackage{algorithm}
\usepackage{algorithmic}
\usepackage{amsmath,bm}
\usepackage[pdftex]{graphicx}
\usepackage{multirow}
\usepackage{array}
\usepackage[justification=centering]{caption}
\usepackage{epstopdf}
\usepackage{comment}
\usepackage{booktabs} 
\usepackage{subfigure}
\newcommand*{\affaddr}[1]{#1}
\newcommand*{\affmark}[1][*]{\textsuperscript{#1}}
\newcommand\blfootnote[1]{%
	\begingroup
	\renewcommand\thefootnote{}\footnote{#1}%
	\addtocounter{footnote}{-1}%
	\endgroup
}

\begin{document}

\title{FacebookVideoLive18: A Live Video Streaming Dataset for Streams Metadata and Online Viewers Locations}

\author{%
	Emna Baccour\affmark[1], Aiman Erbad\affmark[1], Kashif Bilal\affmark[2]\affmark[*], Amr Mohamed\affmark[1], Mohsen Guizani\affmark[1] and Mounir Hamdi\affmark[3].\\
	\affaddr{\affmark[1]CS department, College of Engineering, Qatar University.}\\
	\affaddr{\affmark[2]COMSATS Institute of Information Technology, Pakistan.}\\
	\affaddr{\affmark[3]CSE department, Hamad Bin Khalifa University, Doha, Qatar.}\\
}

\maketitle

\begin{abstract}
With\blfootnote{*This work was done while Dr. Kashif was a visiting research fellow at Qatar University.} the advancement in personal smart devices and pervasive network connectivity, users are no longer passive content consumers, but also contributors in producing new contents. This expansion in live services requires a detailed analysis of broadcasters' and viewers' behavior to maximize users’ Quality of Experience (QoE). In this paper, we present a dataset gathered from one of the popular live streaming platforms: Facebook. In this dataset, we stored more than 1,500,000 live stream records collected in June and July 2018. These data include public live videos from all over the world. However, Facebook live API does not offer the possibility to collect online videos with their fine grained data. The API allows to get the general data of a stream, only if we know its ID (identifier). Therefore, using the live map website provided by Facebook and showing the locations of online streams and locations of viewers, we extracted video IDs and different coordinates along with general metadata. Then, having these IDs and using the API, we can collect the fine grained metadata of public videos that might be useful for the research community. We also present several preliminary analyses to describe and identify the patterns of the streams and viewers. Such fine grained details will enable the multimedia community to recreate real-world scenarios particularly for resource allocation, caching, computation, and transcoding in edge networks. Existing datasets do not provide the locations of the viewers, which limits the efforts made to allocate the multimedia resources as close as possible to viewers and to offer better QoE. 
\end{abstract}
\begin{IEEEkeywords}
Live video streaming dataset, Facebook live, streams pattern, viewers pattern, cloud, edge computing, resource allocation.
\end{IEEEkeywords}
\section{Introduction}
 With the proliferation of live video providers, such as Youtube live, and the availability of pervasive network access, the world has witnessed an explosion in live video streamings. Statistics conducted by Cisco predicted that the mobile videos will present 78\% of the overall mobile traffic by 2021\cite{mobileData}. This evolution is driven further by the advancement of mobile devices that become able to film high quality videos, which encourages the emergence of crowd live streaming. Now, mobile users are no longer passive resource consumers but also contributors to publish video contents. Crowdsourced live videos are now used in a multitude of ways, such as streaming personal events, announcing new products, giving demonstrations, online classes, online gaming, and streaming competitions and contests. One of the most popular live streaming broadcasters is the Facebook platform which has over 2.07 billion monthly active users and 1.15 billion daily \textit{mobile} users \cite{facebook}. According to \cite{facebookstat} and \cite{facebookstat2}, the total views of all shared videos in Facebook was around 32 billion per day in 2016, which means that Facebook can generate over 3,000 years of watch time every day if every view lasts only about 3 seconds. Among these shared videos, 1 out of 5 is live. To keep pace with the unceasing video requests, considerable research efforts are conducted to enhance network capabilities, storage, and computation. 

Multimedia community is endeavoring to maximize the viewers' QoE at minimum cost. Previously collected datasets, such as \cite{twitch2} and \cite{TwitchGame}, were able to provide a glance at the live streams, broadcaster location, and total number of active viewers. However, none of the existing datasets provides the viewers location. Moreover, such datasets are unable to deliver the join and leave time of individual users, neither it is possible to mine the maximum watch time, maximum number of viewers that watched a video for a specified time period, etc. In fact, no detailed information regarding the viewers are available in previous datasets. The absence of fine grained viewers' information may be misleading when considering distributed resource allocation, edge computing perspective, caching, prediction, etc. Therefore, a dataset with detailed and fine grained information about the viewers is required. Unfortunately, none of the live video streaming APIs offer direct fetching of the viewers' statistics. However, Facebook offered little peek to viewers' information by providing a visual display of current live streams and current viewers presented in a live map. Converting exact viewers' locations from a visual map is a tedious task, and the Facebook API does not offer the possibility to extract the set of online videos, the number of viewers and their locations. Therefore, we employed Selenuim webDriver to connect automatically to an active Facebook account and Facebook map. Then, by exploring the HTTP traffic passing through the browser, all online video URLs are captured and from which IDs are extracted. Using these IDs, the Selenuim driver generates the page of viewers' locations related to each video. Using the same method, other metadata can be captured, such as category of the video, broadcaster name and location, broadcasting time, etc. Facebook also provides an API to provide a larger set of metadata for public live videos. Finally, the captured data are stored using MongoDB in a queryable structured format in Json files.

This dataset can be used by the research community to study the challenges faced by live streaming systems and to propose new designs based on real data traces. In this way, the obtained results will have more accuracy and credibility. Similar public datasets such as \cite{twitch2} are collected from Twitch.tv and Youtube Live platforms. These datasets were useful for many research works to test their proposed designs and algorithms. Among them, authors in \cite{Crowdsourced} proposed an optimization framework for resource and computation allocation on cloud. However, since no public dataset includes viewers' locations, the aforementioned work had to add some assumptions related to viewers' positions to obtain a near optimal allocation and transcoding solution. Other efforts such as \cite{kashif2,emna,emna2,emna3, ahmed, vital} designed cloud/edge caching and offloading optimizations, and heuristics aiming to allocate multimedia data in the vicinity of viewers. However, a dataset containing information about requests incoming from small communication areas is not publicly available. This has pushed edge caching authors to use theoretical distributions. 

Based on these challenges, we collected a comprehensive dataset containing thousands of live videos gathered from Facebook in June and July 2018. In addition, the data is collected with many features including creation time, publisher location, and most importantly, locations of viewers with public profiles joining the video during the streaming. The audience positions include the viewers from the same country of the uploader, the viewers in other countries, along with the viewers located in the same neighborhood. Such dataset can help the researchers with various undertakings, such as: (1) a better understanding of the viewers distribution can help to develop  efficient resource allocation algorithms on multi-cloud systems; (2) the variety of features available with the data can help the machine learning community to enhance the allocation and computation models; (3) viewers exhibit different engagement behaviors, which helps to identify which videos to cache or to remove from the edge and also to understand the preference of viewers by region and recommend videos based on these preferences; (4) by studying broadcasters' and viewers' patterns, researchers can develop models to predict required representations and transcoding resources in order to improve viewers' QoE. To the best of our knowledge, there is no existing public dataset that tracks the locations and behaviors of live viewers along with the characteristics of videos. The contribution of our dataset not only lies in covering all published live videos during a specific period, but also the diversity of video types, countries, and locations. The contributions in this paper are summarized as follows:
\begin{itemize}
    \item We present a dataset of traces from Facebook, collected in June and July. Our dataset contains more than 1,500,000 live streams, available on our public website\footnote{https://sites.google.com/view/facebookvideoslive18/home}.
    This dataset can provide the basis for multiple researches either on live streaming systems or on cloud and edge infrastructures.
    \item We explain the steps and the challenges faced to collect the data and we describe the structure of the dataset.
    \item We show preliminary studies on the dataset. Specifically, we describe the status of the system including the number of broadcasted videos, the number of broadcasters, the geographical distribution of streams, the evolution of streams over time (hours and days), the broadcasting behavior of streamers, etc. We show that streams follow a similar pattern during different days of the week, and we observe that the list of loyal uploaders for live activity can be defined.
    \item We also go deeper by studying the viewing behavior, the popularity of videos, the engagement of viewers and their geographical distribution. We confirm that the distribution of views follows Pareto law. More interestingly, we highlight that most of the viewers are located in the same area of the content broadcaster.
\end{itemize}

The rest of this paper is organized as follows: Section \ref{Related_work} describes some related works. Section \ref{Data_Collection} presents the dataset and the data collection methodology. Section \ref{isight} provides some of our introductory analyses on the dataset. Section \ref{application} gives an example of the usage of our dataset. Finally, we conclude in section \ref{conclusion}.
\section{Related work}\label{Related_work}
Many efforts have been conducted to capture datasets targeting generated content live streaming. Most of these efforts are dealing with the game-casting, which means the gamers that are broadcasting their activities during a game. Collecting the meta-gaming is called "the game beyond the game", which refers to the connections between the broadcasters and their viewers, the popularity of channels, the peak hours of games and viewing, the most viewed games, etc. These meta-game can help the gamers to rethink the way, the time, and the games they are playing, and can guide the sponsors and game publishers to rethink the strategy of advertising. Recently, such data has been also used by network engineers to optimize the allocation of videos on the cloud and enhance the quality of service experienced by the viewers.

Authors in paper \cite{onlineGames} studied the characteristics of game-casting traffic for different game types including real-time strategy (RTS) games, card and board games, and sports games broadcasted. This study found that the games popularity follows a power-law and the games workloads collected from all types of games are similar and predictable in a short time interval. In addition, the players' churn increases over time and the play behavior changes when the gamers are about to leave together.  Authors in \cite{XFire} collected traces from XFire gaming platform. XFire is a social website that has attracted 25 million users sharing their games activities. The authors collected the  traces tracking games of 20 million users playing over 1,500 games. Then, they proposed a method to study the characteristics of the online meta-gaming networks (OMGN) and observed several facts: First, the OMGN players spend in-game over 100 years hourly, a significant percentage of players played over 10,000 hours of games, finally, users of the platform are routinely engaged in sharing their videos. A more recent study was conducted by authors in \cite{twitch2}. The paper presents a dataset captured  from the popular gaming platform Twitch.tv during 3 months in 2014 with a broad population of uploaders. This dataset includes also traces captured from Youtube live platform. The total number of viewers, the session popularity, the total number of concurrent sessions,  and some channels metadata are also fetched. 

None of the aforementioned studies have collected a data related to viewers' locations and community preferences. The main novelty of our paper compared to previous works is that it presents the distribution of viewers during the streaming along with richer metadata.
\section{Data Collection}\label{Data_Collection}
\begin{figure}[!h]
	\centering
	\includegraphics[scale=0.2]{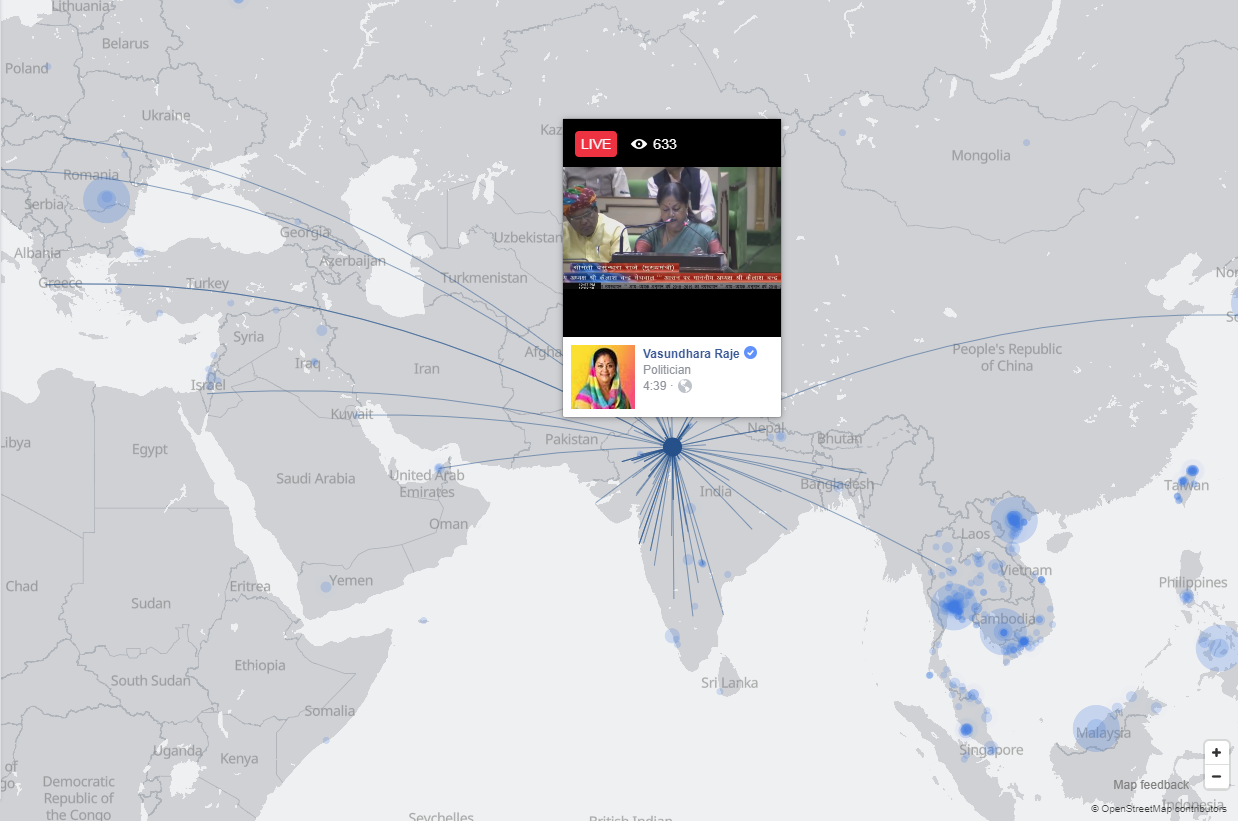}
	\caption{Facebook live map.}
	\label{live}
\end{figure}

Facebook Live Map\footnote{https://www.facebook.com/livemap/} is a visual display of current live streams and current viewers. Our objective is to
explore the data behind this map starting from different live streams, their related metadata, and the
viewers of these videos. In fact, Facebook Live Map is a dynamic platform. Each new public stream
appears on the map as a dot representing the video uploader's location. The dot gets larger as viewers
join in and the overall viewers count increases. When we hover over these dots, one can see lines
extending outwards that represent the physical locations of the viewers as illustrated in Figure \ref{live}. These
locations are only displayed when a video reaches 100 views count. The live videos presented in the
map are those published by public pages or profiles and the locations are those of viewers with no
privacy settings. Facebook does not provide an API to collect the IDs of videos and the locations of
viewers. Therefore, we developed a set of synchronized Python scripts to capture, from the map, a
global view of the real time streaming system, in different time instants. In fact, in each fetch, the
collection of identifiers (IDs) of the ongoing videos is not directly possible from the map. Yet, we can
extract the post page of each video, from where we can capture the ID of the online stream. Similarly,
the current number of viewers, and the locations of broadcasters can also be collected. After gathering different IDs, we explored the HTTP traffic passed through the browser to find the URL serving to display
the longitude and latitude of different viewers. Unfortunately, the live map website is no longer
available and the described display is shut down by Facebook, which makes our dataset more valuable.

Facebook also provides a Graph API that enables to get the metadata of video streams at any time,
when having their IDs. Hence, after the live streams are captured from the map, their IDs are parsed to
get the metadata of different videos and their publishing pages using Facebook Graph API. Information
of personal profiles and videos are not included in the public dataset to protect users' privacy. The
information fetched from Facebook Graph API are richer than those provided with Twitch and Youtube
Live dataset described in \cite{twitch2}. To the best of our knowledge, we are the first to track the viewers’
locations and the number of views with a small interval granularity.
We used various programming languages and database tools to extract, store, and parse the data
including Python, MongoDB, and Selenuim webDriver. The database presented as Json files and the
fetching and parsing scripts are available on our public website.
\section{A first insight into the data}\label{isight}
In this section, we present our dataset and analyze its characteristics that can provide us with a first
insight into the Facebook live community, as well as worldwide spectators. In this paper, we will
conduct our studies on data collected in a period from June 3$^{rd}$, 2018 to July 6$^{th}$, 2018. As a result, we
obtained a list of 1.506.473 videos. Table \ref{characteristics} gives a summary of this dataset on which we conducted our
analyses.
\begin{table}[h]
	\centering
	\caption{Summary of the dataset.}
	\label{characteristics}
	\begin{tabular}{|l|l|}
		\hline
		\multicolumn{1}{|c|}{Period} & June 3, 18- July 6, 18 \\ \hline
		\multicolumn{1}{|c|}{\begin{tabular}[c]{@{}c@{}}Number of videos\end{tabular}} & 1.506.473 \\ \hline
		\begin{tabular}[c]{@{}l@{}}Total number of broadcasters\end{tabular} & 408.231 \\ \hline
		\begin{tabular}[c]{@{}l@{}}Number of  categories\end{tabular} & 1.151 \\ \hline
		viewers\textgreater100 &  18,34 \%\\ \hline
	\end{tabular}
\end{table}

We observe in our measurements that the videos are classified into 1.151 categories and published by
408.231 different broadcasters. The categories of videos include news, public figures, sports, music, etc.
However, the top five most popular categories are \textit{"Media/News Company"} with 52.210 videos,
\textit{"Broadcasting \& Media Production Company"} publishing 35.091 live streams, \textit{"Radio Station"} having
24.897 videos, \textit{"Religious Organization"} and \textit{"TV Channel"} publishing 24.843 and 23.344 videos,
respectively. Also, we notice that only 18,43 \% of videos have more than 100 simultaneous viewers
while being live. This can be explained by the fact that most of the videos are captured by users’ profiles
that do not have a large number of friends. 

\begin{figure}[!h]
	\centering
	\includegraphics[scale=0.7]{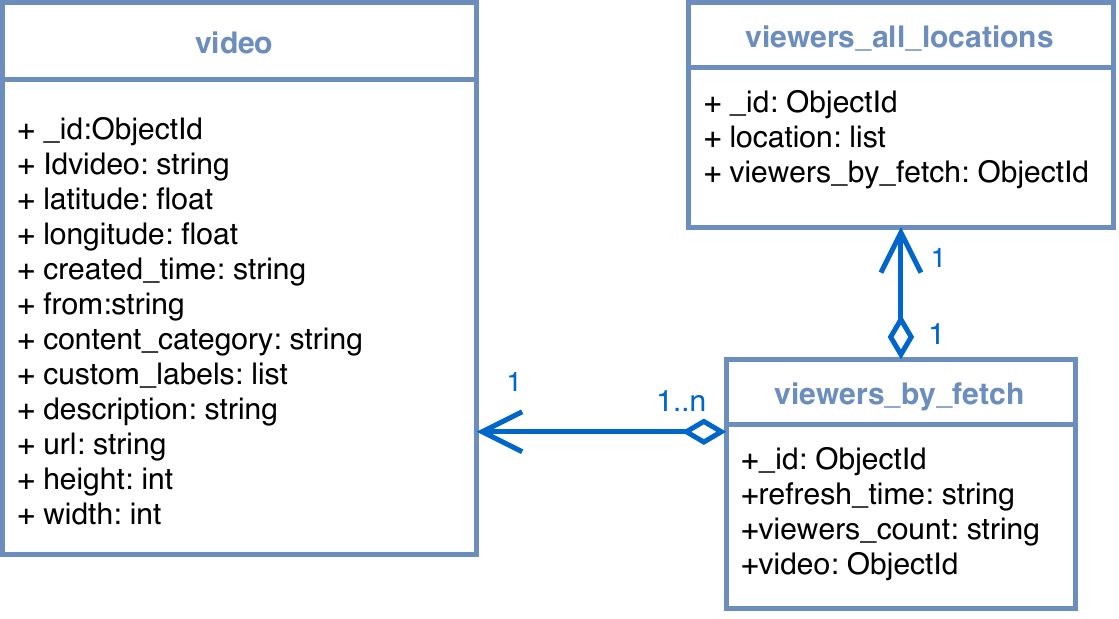}
	\caption{Structure of the dataset.}
	\label{dataset1}
\end{figure}

\begin{figure}[!h]
	\centering
	\hspace{-0.6cm}
	\includegraphics[scale=0.39]{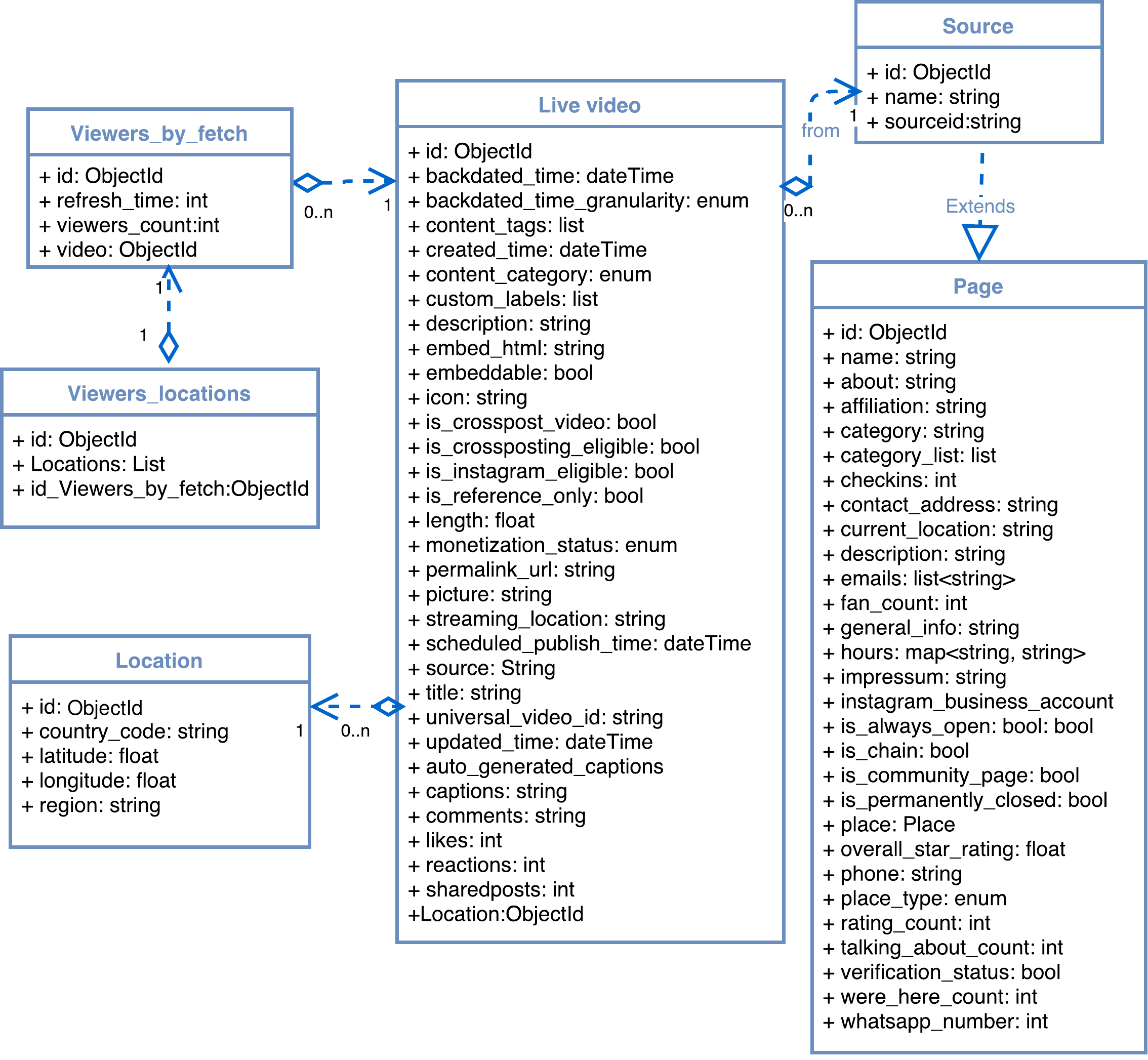}
	\caption{Structure of the extended dataset.}
	\label{dataset}
\end{figure}

The structure of our dataset is presented in Figure \ref{dataset1}.
Each video class is defined by its dataset \textit{"\_id"}, the identifier given by Facebook \textit{"idvideo"}, the latitude and
longitude of the broadcaster location, the broadcasting time \textit{"created\_time"}, the broadcaster name
\textit{"from"}, the content category, the URL of the video, and finally, the height and the width of the content,
which serve to determine the quality of the video. Since we fetch the data from the map every interval
of time, a video can be related to multiple fetches. Each fetch is defined by an \textit{"\_id"}, the time of the fetch
\textit{"refresh\_time"}, the number of viewers at the time of the fetch and the related video id. For each fetch, a
list of viewers’ locations is created, namely \textit{"viewers\_all\_locations"}. Other metadata can be captured
using the Facebook API to create a more complete dataset, as illustrated in Figure \ref{dataset}. This API allows
capturing more details about the video. We created this complete dataset for a data collected in January
and February\footnote{https://sites.google.com/view/facebookvideoslive18/home}. The code to extend the dataset of June and July discussed in this paper is provided in our
dataset website\footnote{https://sites.google.com/view/facebookvideoslive18/get-your-dataset?authuser=0}. 
\subsection{Geographic distribution of live videos}\label{Geographic_distribution}
\begin{figure}[!h]
	\centering
	\mbox{
		\hspace{-0.4cm}
		\subfigure[\label{countries1}]{\includegraphics[scale=0.34]{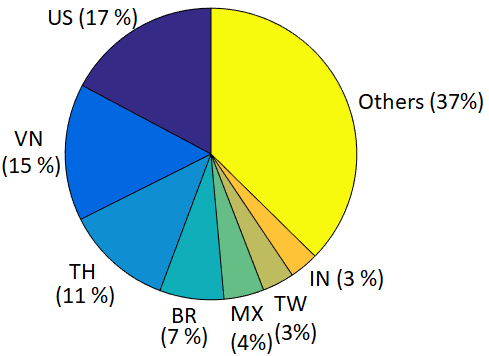}}
			\vspace{-0.3cm}
		\subfigure[\label{countries2}]{\includegraphics[scale=0.34]{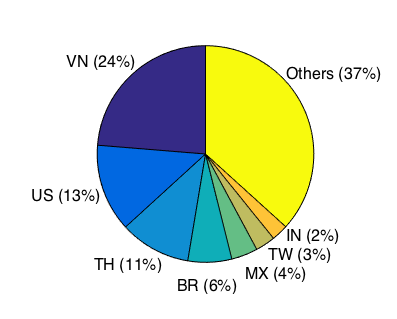}}}

	\caption{Geographical distribution of live streams: (a) Geographical distribution of  live streams in Feb 9$^{th}$,18 – Feb 16$^{th}$,18 (b) Geographical distribution of live streams in June 3$^{rd}$,18 – June 9$^{th}$,18.}
	\label{map}
\end{figure}

The geographical distribution of Facebook live videos is illustrated in Figure \ref{map}. The data pattern follows our expectations, showing that the United States has the highest activity in February 2018, which is expected since Facebook is the most known social media in the USA (Figure \ref{countries1}). In June, the number of videos broadcasted from Vietnam became larger than those broadcasted from the United States (Figure \ref{countries2}). This can be explained by the Cambridge Analytica data suit faced by Facebook, in March 2018. After that scandal, American citizens started to become more aware of data leaking and to convert their profiles and publications to private. Hence, the number of collected data from the US was scarcer after March.
\begin{figure*}[!h]
	\centering
	\mbox{
	    \hspace{-0.5cm}
		\subfigure[\label{long1}]{\includegraphics[scale=0.37]{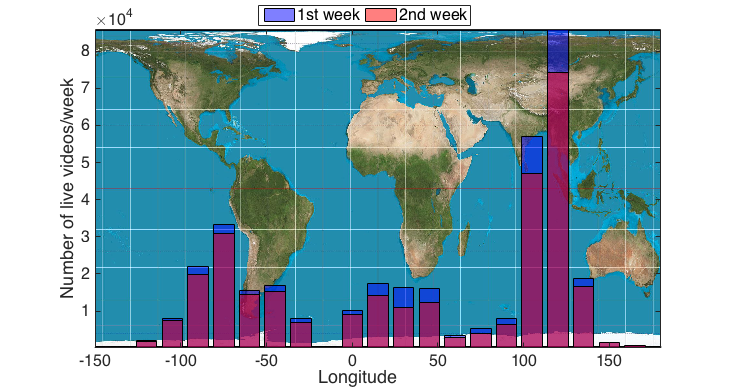}}
		\hspace{-0.3cm}
		\subfigure[\label{long2}]{\includegraphics[scale=0.37]{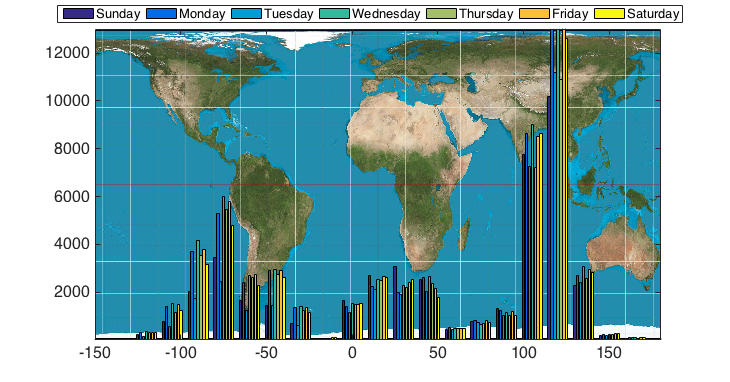}}}

	\caption{Longitudinal distribution of live streams: (a) longitudinal distribution of live videos during the first two weeks of June: The Figure shows that the live streaming follows almost the same pattern during the two studied weeks (b) longitudinal distribution of live videos during different days of the first week of June: The Figure shows that each longitudinal region is characterized by a similar broadcasting pattern in different week days.}
	\label{map2}
\end{figure*}

Figure \ref{map2} presents per-longitude histograms. Figure \ref{long1} presents the number of live videos occurring in different longitude regions during the first two weeks of June. We can see that most of the live videos originate from Asia (particularly Vietnam (VN) and Thailand (TH)) and North and South America. Additionally, we can notice that the number of captured videos is almost similar between the two weeks for most of the regions (The red and blue graphs are superposed). Figure \ref{long2} shows the number of videos published on different days of the first week of June and distributed in different longitude regions. We observe that each longitudinal region is characterized by a similar broadcasting pattern on different days of the week. Note that the fetching time follows UTC+3 timezone.
\subsection{Stream and Streamer characteristics}
\begin{figure}[!h]
	\centering
	\includegraphics[scale=0.5]{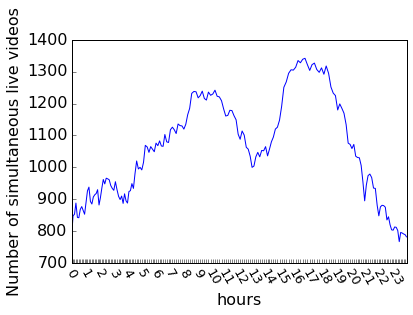}
	\caption{Number of simultaneous live videos/day.}
	\label{videos_hours}
\end{figure}

\begin{figure}[!h]
	\centering
	\includegraphics[scale=0.5]{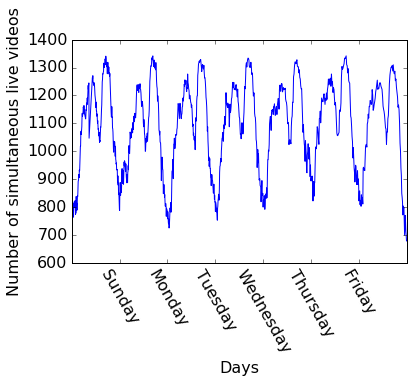}
	\caption{Number of simultaneous live videos at different days of the week.}
	\label{videos_days}
\end{figure}

The streaming activity depends on different parameters, including the popularity of the social media in different countries/timezones and the activity of the uploaders. We measure the number of simultaneous ongoing videos from the collected data in different instants of the day to analyze the daily and weekly patterns of the live activity. Figure \ref{videos_hours} presents the number of ongoing streams during one day. It clearly appears that the live broadcasting has two peak periods, which are the interval from 5 to 13 and the interval from 14 to 21 (UTC+3). These periods correspond to the daytimes of Vietnam and east USA. Moreover, the valley period from 21 to 4 corresponds to the night time of Vietnam zone broadcasting the quarter of videos. These facts are expected since we showed in section \ref{Geographic_distribution} that Facebook live activity is popular in America and Vietnam zones. Figure \ref{videos_days} presents the number of active streams on each day of the week. We can see that the publishing activity follows the same pattern for different days, which shows that our data is valid for machine learning tasks.

\begin{figure}[h]
	\centering
	\mbox{
		\hspace{-0.8cm}
		\subfigure[\label{videos_broadcasters}]{\includegraphics[scale=0.32]{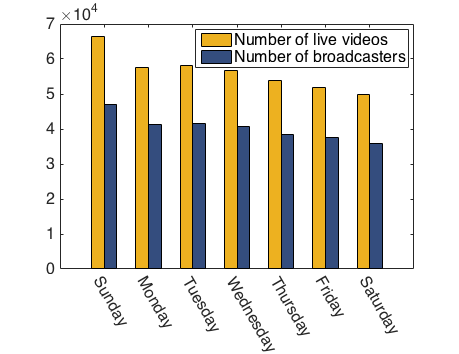}}\hspace{-0.5cm}
		\subfigure[\label{videos_broadcaster}]{\includegraphics[scale=0.32]{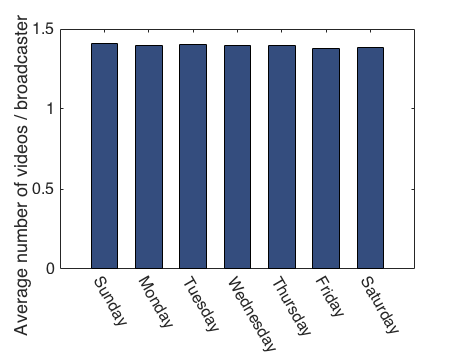}}
	}
	\caption{Number of videos VS number of broadcasters: (a) shows that many publishers upload more than one video per day. (b) shows that Facebook live service has a near-constant proportionality between videos and their publishers.}
	\label{videos_vs_broadcasters}
\end{figure}

Figure \ref{videos_broadcasters} shows the number of broadcasted videos compared to the number of publishers among
different days of the first week of June. We can see that many publishers upload more than one video
per day. Figure \ref{videos_broadcaster} presents the average number of published live videos per broadcaster and the
proportionality between streams and streamers during different days of the week. It can be seen that
Facebook live service has a near-constant proportion between videos and their publishers.
For better understanding, we try to estimate the loyalty of uploaders to the broadcasting activity.
Hence, we calculated the number of videos published by each channel. We found that, on average,
around 43\% of users publish 2 or more videos per day and only 15\% of users publish more than 5 videos
per week. Among the most active broadcasters, we can cite the Tv news channel \textit{"BtvNews"} publishing
868 videos per month. 

After analyzing the streams' and streamers' characteristics, we can notice that
streams follow an almost similar pattern among different days of the week and that we can derive the
set of the most loyal uploaders to live activity. These facts can encourage the machine learning
community to study the user-generated live streaming systems.
\begin{figure}[!h]
	\centering
	\includegraphics[scale=0.48]{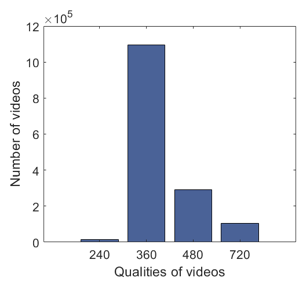}
	\caption{Distribution of videos’ qualities.}
	\label{bitrate1}
\end{figure}
 
Next, we present in Figure \ref{bitrate1} the distribution of videos’ qualities published by different broadcasters.
We can see that most of the videos are broadcasted with a low quality (360p). This can be explained by the
fact that more than 40\% of live videos are published from Asia and that 21\% of screen resolutions of
broadcasters’ phones in this continent do not allow to capture video qualities higher than 360p \cite{screen}. Note
that the highest bitrate representation allowed by Facebook to broadcast live videos is 720p.
\begin{figure}[!h]
	\centering
	\includegraphics[scale=0.5]{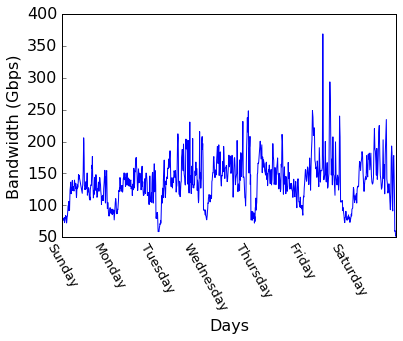}
	\caption{Bandwidth usage for live video delivery.}
	\label{bitrate2}
\end{figure}

We then evaluate the overall bandwidth used by the live streaming system. We consider that Facebook
is an Over-The-Top (OTT) service with unicast transmission to viewers, so we sum the bitrates of each
session and we multiply it by the number of viewers for this stream. We set the bitrate variants related
to each representation to be 0.45, 0.55, 0.67, 0.82 of the original video bitrate version and we consider
that all videos have the same original bitrate equal to 2 Mbps. Figure \ref{bitrate2} shows that Facebook live
service experiences a  bandwidth peak of more than 350 Gbps to deliver only public videos. Such a
volume of bandwidth matters not only for the live streaming services, but also for operators, as they
need to deliver this huge amount of information to end users. We note that the contents are live, and
therefore cannot be pre-fetched or cached.
\subsection{views and viewers characteristics}
\begin{figure}[!h]
	\centering
	\includegraphics[scale=0.5]{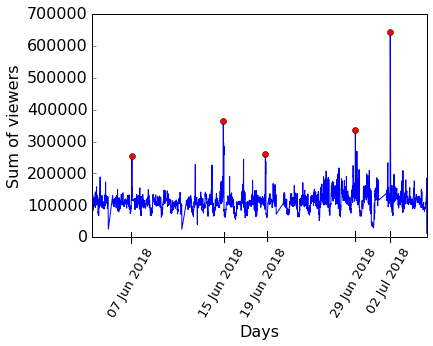}
	\caption{Cumulative number of viewers at different days: Instants witnessing peak viewing are marked in red.}
	\label{viewers_days2}
\end{figure}

We describe in this section the characteristics of viewers and the distribution of their locations. These
characteristics distinguish our dataset from other datasets missing this information. Figure \ref{viewers_days2} shows the
cumulative number of views among different days of the studied period (June and July). In this figure,
we marked the instants witnessing a peak number of viewers. These viewing peaks are related to the
broadcasting of special events. For example, on June 7$^{th}$, the 2018 NBA final was broadcasted. On June
15$^{th}$ and June 19$^{th}$, there were two important football matches; which are respectively Spain vs Portugal and
Egypt vs Russia 2018 World Cup matches. Finally, June 29$^{th}$ and July 2$^{nd}$ correspond to the rescue days of Thailand’s cave accident.

\begin{figure}[h]
	\centering
		\subfigure[\label{video_popularity}]{\includegraphics[scale=0.4]{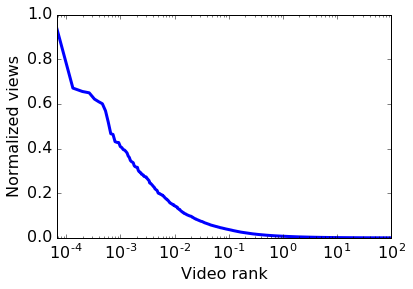}}\\
		\subfigure[\label{views_track}]{\includegraphics[scale=0.4]{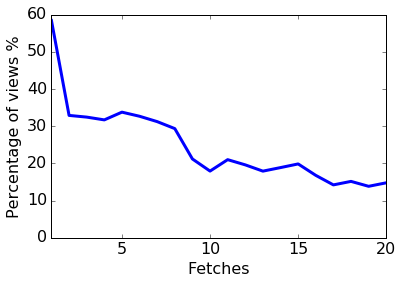}}		
	\caption{Video popularity: (a) Facebook live system follows Pareto Principle. (b) The engagement of viewers starts dropping after a period of streaming.}
	\label{popularity}
\end{figure}

We also present the popularity of live sessions, which we define in this paper as the number of views
per video. Since the number of viewers during an ongoing video is not constant, we define the video
popularity as the peak number of its views during the collection time. The average popularity of all video
sessions is 99 views. However, as illustrated in Table \ref{characteristics}, only 18\% of videos have more than 100 viewers,
which means that most of the streams attract a very small audience compared to the popular streams
that attract a huge number of viewers. In particular, 0.04\% of videos were viewed by more than 10,000
simultaneous viewers, 1.15\% were viewed by more than 1000 simultaneous viewers and 18.4\% of
videos were viewed by more than 100 simultaneous viewers. The average popularity of this 18\% most
viewed videos is equal to 410 views. To deeply evaluate the videos' popularity distribution, we verify, in
the following, if the Facebook live system follows Pareto Principle. Figure \ref{video_popularity} illustrates the popularity of
different videos in the studied period. Videos ranks are normalized between 0 and 100 and views are
normalized between 0 and 1. We can see that more than 90\% of views are related to the top 10\% of
videos, which follows Pareto law. Similar analyses of popularity were made for Twitch \cite{videoGaming} and proved that 80\% of views account for the top 10\% of popular videos. 

The videos' popularity is also evaluated by studying the engagement of viewers, which means how long
the video attracts viewers and maintains an increasing viewing pattern. Figure \ref{views_track} presents
the viewers engagement to videos equal to or longer than 1 hour. We follow the activity of viewers every
3 minutes for 1 hour, then, we measure the percentage of viewers in each fetch compared to the total
number of views. We can see that the number of views is at its maximum in the beginning of the
streaming. However, the engagement of viewers starts dropping after a period of streaming, which is equal to 2
fetches (6 minutes) on average as shown in Figure \ref{views_track}. The number of views drops again after 8 fetches
(24 minutes) and the viewers that are loyal to the video will likely stay engaged to it.

\begin{figure}[!h]
	\centering
		\subfigure[\label{video_locations}]{\includegraphics[scale=0.4]{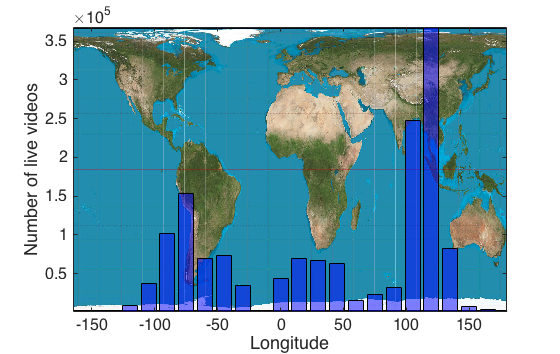}}
		\\
		\subfigure[\label{viewers_locations}]{\includegraphics[scale=0.4]{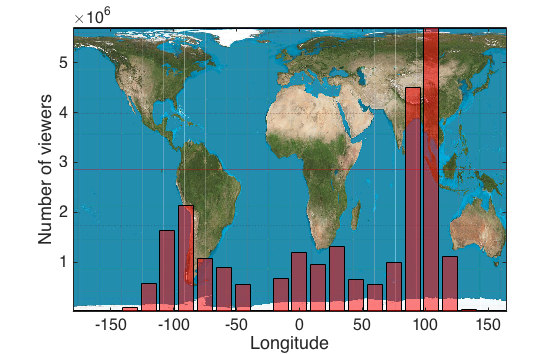}}
	
	\caption{Locations of broadcasters VS locations of viewers.}
	\label{popularity}
\end{figure}

Compared to other datasets, the competitive edge of our dataset is the tracking of viewers’ locations in
each fetch. The fetched locations include only the viewers with public profiles. Hence, the list of
locations will not contain all audience positions. To estimate the complete list of all current viewers, we
applied the statistical inference theory with 90\% confidence. Figure \ref{video_locations} presents the per-longitude
histogram of all the broadcasters' locations. Figure \ref{viewers_locations} presents the viewers locations during the
collection period. We can see that viewers originate from the same time zone of the uploader. This fact
can help the edge computing community to enhance their data allocation systems.

\section{Applications}\label{application}
Our FacebookVideoLive18 dataset can be used to study different challenges faced in large live streaming systems. In fact, a researcher can recreate the live sessions scenarios and evaluate the results based on real data traces, which adds more credibility and value to one's work. One of the possible usages of the data is the multi-cloud resource allocation. In fact, in live streaming systems, the crowdsourcers upload their videos in real time. These videos will be viewed by lively high distributed viewers. Broadcasters and viewers are heterogeneous in terms of locations, network bandwidth, hardware configurations, etc. Due to this heterogeneity and the increasing demand for live services, an immense amount of cloud storage and transcoding resources is required. In addition, more stringent delay constraints are imposed, which makes the problem more challenging. Many efforts, such as \cite{Crowdsourced}, have proposed geo-distributed cloud designs to solve the resource allocation problems and push the multimedia data as close as possible to the viewers. Publicly available real-world traces do not include the locations of viewers, which limits the efficiency of resource allocation models. Our data will enable multi-cloud community to predict the locations and requirements of viewers with higher credibility. Also, due to the diversity of features and metadata presented with the traces, machine learning techniques can be adopted to estimate the required computing and storage resources and find the optimal allocation solution.   

Another potential usage of the data is the edge caching and traffic allocation problem. Indeed, one of the main challenges is to reduce as much as possible the data exchange between end-users and remote clouds. Therefore, the research community has become more interested in edge solutions and crowdsourced caching and offloading. However, because of the unavailability of requests' and viewers' locations, researchers are using different theoretical distributions \cite{emna2,emna3}. According to our data containing the audience positions, we proved that most of the viewers are located in the proximity of the video uploader. Also, since some viewers are located in the same neighborhood, D2D caching and offloading systems can be evaluated using our real-world data. An Examples of how our dataset can be interpreted and used are found in our research articles \textit{“QoE-Aware Resource Allocation for Crowdsourced Live Streaming: A Machine Learning Approach”} \cite{Fatima} and \textit{Transcoding Resources Forecasting and Reservation for Crowdsourced Live Streaming} \cite{globecom}.

\section{Conclusion}\label{conclusion}
In this paper, we have presented a dataset of Facebook live streaming videos. We have shown that most of the videos during the studied period originated from east Asia. Moreover, we have noticed that broadcasting patterns remain alike in all days of the week. In addition, we have proved that the videos popularities follow Pareto law and that a live video has a descending viewing pattern after a streaming time. Finally, it has been observed that most of the viewers are located within the same region as the video uploader. This dataset can help the research community to study and understand the behavior of the crowdsourcing live content systems. To open new directions, we have highlighted some problems and research areas that can be well studied by using our real data traces. We hope that FacebookVideoLive18 helps the multimedia community to understand the characteristics of their systems and improve them accordingly. 
\section*{Acknowledgement}
This publication was made possible by NPRP grant 8-519-1-108 from the Qatar National Research Fund (a member of Qatar Foundation). The findings achieved herein are solely the responsibility of the author(s).
\bibliographystyle{IEEEtran}
\bibliography{db}
\end{document}